\documentclass[aps,prb,reprint,amsmath,amssymb,a4paper,floatfix,showpacs]{revtex4-1}
\usepackage{mathptmx}
\usepackage[utf8]{inputenc} 
\usepackage{graphicx}
\usepackage{xspace}
\usepackage[
,textwidth=17.4cm
,textheight=25.3cm
,verbose
,dvips
]{geometry}

\begin{document}

\title{Coexistence of quantum-confined Stark effect and localized states\\ in an (In,Ga)N/GaN nanowire heterostructure}

\author{Jonas Lähnemann}
\email{laehnemann@pdi-berlin.de}
\author{Oliver Brandt}
\author{Carsten Pfüller}
\author{Timur Flissikowski}
\author{Uwe Jahn}
\author{Esperanza Luna}
\author{Michael Hanke}
\author{Matthias Knelangen}
\author{Achim Trampert}
\author{Holger T. Grahn}
\affiliation{Paul-Drude-Institut für Festkörperelektronik,
Hausvogteiplatz 5--7, 10117 Berlin, Germany}

\begin{abstract} We analyze the emission of single GaN nanowires with (In,Ga)N insertions using both micro-photoluminescence and cathodoluminescence spectroscopy. The emission spectra are dominated by a green luminescence band that is strongly blueshifted with increasing excitation density. In conjunction with finite-element simulations of the structure to obtain the piezoelectric polarization, these results demonstrate that our (In,Ga)N/GaN nanowire heterostructures are subject to the quantum-confined Stark effect. Additional sharp peaks in the spectra, which do not shift with excitation density, are attributed to emission from localized states created by compositional fluctuations in the ternary (In,Ga)N alloy.
\end{abstract}

\pacs{78.67.Uh,%Optical Properties of Nanowires 
73.22.-f,%Electronic structure of nanoscale materials and related systems 
78.55.Cr,%PL; III-V semiconductors
78.60.Hk%,%Cathodoluminescence, ionoluminescence 
%78.47.jd,%Time resolved luminescence
%71.55.Eq,%Impurity and Defect Levels; III-V semiconductors 
%73.63.Hs,Electronic transport in nanoscale materials and structures; Quantum wells 
%68.37.Lp,%Transmission electron microscopy (TEM) 
%68.37.Og,%High-resolution transmission electron microscopy (HRTEM) 
%68.37.Hk,%Scanning electron microscopy (SEM) (including EBIC) 
}

\maketitle

\section{Introduction}

(In,Ga)N/GaN nanowire (NW) heterostructures have the potential to improve optoelectronics in the visible spectral range.\cite{Kikuchi_jjap_2004} They offer the possibility for dislocation- and strain-free group-III-nitride growth on economically attractive Si substrates.\cite{Calleja_prb_2000,Lagally_nat_2004,Brandt_prb_2010} Several groups have recently reported the growth of (In,Ga)N insertions (axial heterostructures) in NWs.\cite{Kikuchi_jjap_2004,Kishino_procspie_2007,Armitage_mrssp_2009,Chang_apl_2010,Sekiguchi_apl_2010,Bavencove_pssa_2010,Armitage_nt_2010,Lin_apl_2010,Nguyen_nl_2011,Knelangen_nt_2011} An absence of the quantum-confined Stark effect\cite{Miller_prl_1984,Takeuchi_jjap_1997} (QCSE) is reported in some of these studies,\cite{Bardoux_prb_2009,Armitage_nt_2010,Lin_apl_2010,Nguyen_nl_2011} and it is concluded that the internal piezoelectric fields in the (In,Ga)N insertion are ``not significant''.\cite{Bardoux_prb_2009,Lin_apl_2010,Nguyen_nl_2011} As a possible origin for such a drastic reduction in the piezoelectric polarization, the efficient strain relaxation in the NW geometry is suggested.\cite{Bardoux_prb_2009,Lin_apl_2010} 

The emission from a NW ensemble represents a superposition of spectra from single NWs. To truly understand the origin of the observed luminescence, it is necessary to examine the emission of individual NWs. Only few reports of such measurements for (In,Ga)N/GaN NW heterostructures have been presented so far.\cite{Kawakami_apl_2006,Bardoux_prb_2009} \citet{Bardoux_prb_2009} reported an emission they attributed to localized states. In conclusion to their findings, they stress the necessity for a detailed strain calculation in conjunction with a precise structural analysis. Such a study has only recently been presented in Ref.~\onlinecite{Knelangen_nt_2011}. 

In the present paper, we report on both cathodoluminescence (CL) and micro-photoluminescence ($\mu$PL) measurements on single NWs from the sample characterized in Ref.~\onlinecite{Knelangen_nt_2011} to learn more about the mechanism governing the observed green luminescence band. We find conclusive evidence that, in addition to the localization of excitons, polarization fields still play an important role in these (In,Ga)N/GaN NW heterostructures. 

\section{Experimental}

The (In,Ga)N/GaN NW heterostructure was grown in a self-induced approach using plasma-assisted molecular-beam epitaxy on Si(111) substrates.\cite{Knelangen_nt_2011} Its structure and composition were deduced in Ref.~\onlinecite{Knelangen_nt_2011} from a careful geometrical phase analysis (GPA) of high-resolution transmission electron micrographs and from synchrotron-based high-resolution x-ray diffraction (XRD) scans. The two In$_x$Ga$_{1-x}$N insertions of height $h \approx 11$~nm are separated by a 2--3~nm wide barrier and fully embedded in the GaN matrix. The insertions have an average In content of $x \approx 0.2$, but segregation of In resulted in a 1--2~nm thick cap with $x \approx 0.4$. The broad room-temperature PL band in the green spectral region was ascribed to an inter-well transition between the two (In,Ga)N insertions.\cite{Knelangen_nt_2011}

For the optical analysis of single NWs, we utilize both a Gatan Mono-CL3 CL system and a Jobin-Yvon $\mu$PL setup. The CL system is equipped with a photomultiplier and a charge-coupled device (CCD) detector mounted to a Zeiss Ultra55 field-emission scanning electron microscope (SEM).\cite{Jahn_prb_2010} The measurements are performed with a spectral resolution of 10~meV. The $\mu$PL setup uses the 325~nm line of a Kimmon HeCd laser for excitation focused to a spot diameter of about 3~$\mu$m,\cite{Pfuller_prb_2010} a spectral resolution set to 1~meV and a CCD for detection. In either system, the samples can be cooled with liquid He to 10~K. For single NW spectroscopy and the correlation with transmission electron microscopy (TEM), the as-grown NW ensemble is transferred either mechanically to a gold-coated Si(111) substrate or ultrasonically to 2-propanol and dispersed onto a TEM finder grid. The grid is supported by a carbon film for TEM, while it is placed on a Si(111) substrate sputter-coated with a few nm of Au for spectroscopy. Time-resolved PL (TRPL) is carried out on the NW ensemble at both 20~K and 300~K using a spectrometer with 5~meV spectral resolution together with a streak camera for detection. The sample is excited by laser pulses with a wavelength of 389~nm and 200~fs duration provided by the second harmonic of a Coherent Mira~900 oscillator. The repetition rate was set to 603~kHz using a pulse picker.

%=====================================================================
%%Fig.1
\begin{figure}
\includegraphics*[width=6.9cm]{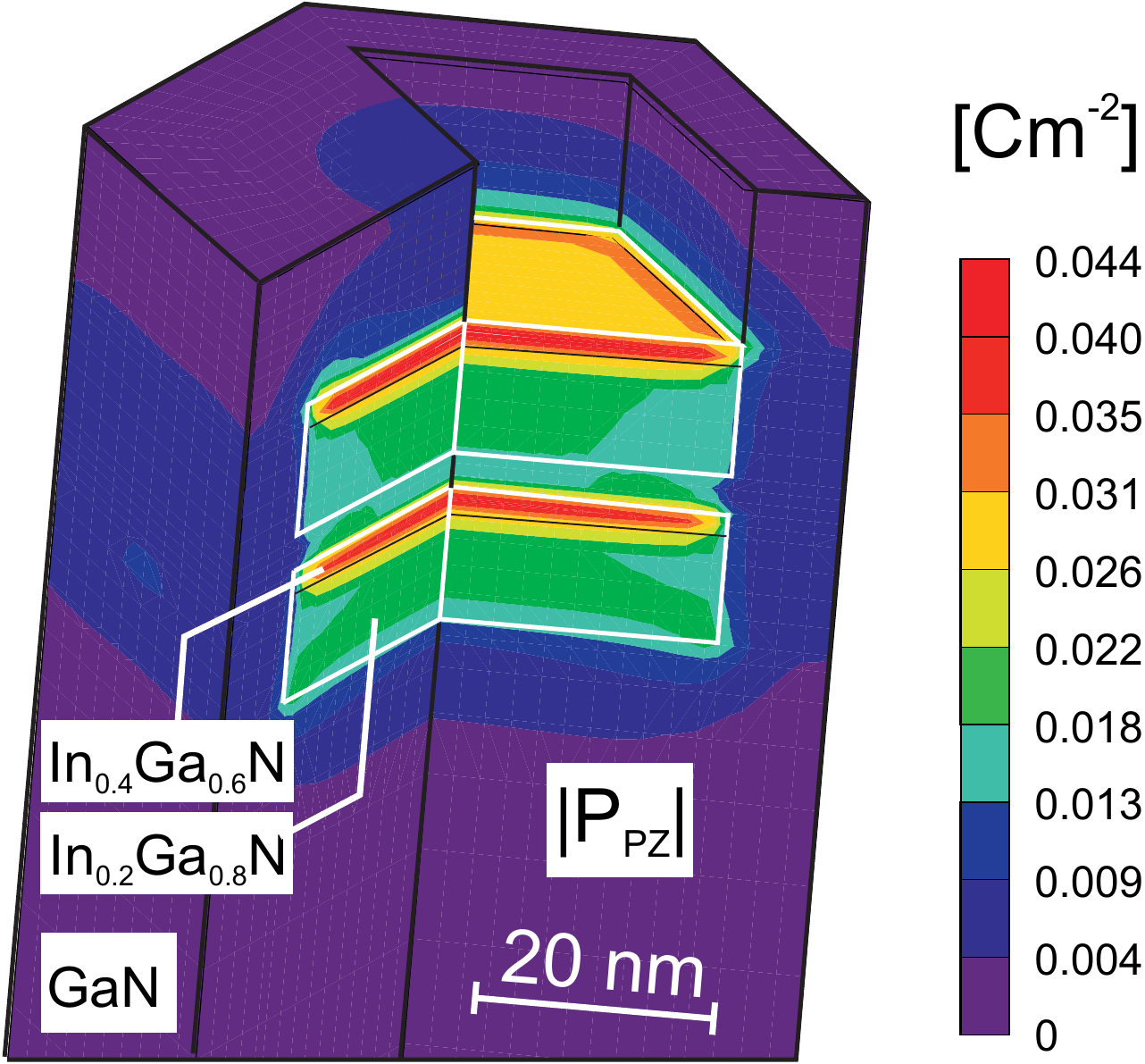} 
\caption{\label{fig:FEM}(Color online) Spatial distribution of the absolute value of the piezoelectric polarization within the NW heterostructure as obtained by FEM simulations. The geometry and composition of the structure has been derived by TEM and XRD.\cite{Knelangen_nt_2011}}
\end{figure}
%=====================================================================

\section{Results and discussion}

Figure~\ref{fig:FEM} visualizes the structure described above and shows the spatial distribution of $|\vec{P}_{PZ}|$, the absolute value of the piezoelectric polarization, which is the dominant cause of the QCSE in (In,Ga)N/GaN heterostructures. The elastic strain tensor $\epsilon_{ij}$ was obtained by numerical finite-element (FEM) simulations based on linear elasticity theory\cite{Christiansen_pssa_1996} with the structural parameters from TEM and XRD as input parameters. The piezoelectric polarization $\vec{P}_{PZ}$ is related to $\epsilon_{ij}$ via the piezoelectric tensor for the spacegroup P6$_3$mc: 
\begin{equation}
\vec{P}_{PZ}=
\begin{pmatrix}
0 & 0 & 0 & 0 & e_{15} & 0 \\
0 & 0 & 0 & e_{15} & 0 & 0 \\
e_{31} & e_{31} & e_{33} & 0 & 0 & 0 
\end{pmatrix} 
\begin{pmatrix}
\epsilon_{xx} \\
\epsilon_{yy} \\
\epsilon_{zz} \\
\epsilon_{yz} \\
\epsilon_{xz} \\
\epsilon_{xy}  
\end{pmatrix}
\label{PiezopolEq}
\end{equation}

The quantity displayed in Fig.~\ref{fig:FEM} is the absolute value of $\vec{P}_{PZ}$, i.e., 
\begin{equation}
|\vec{P}_{PZ}|=
\begin{vmatrix} 
e_{15}\epsilon_{xz} \\
e_{15}\epsilon_{yz}  \\ 
 e_{31}\left(\epsilon_{xx}+\epsilon_{yy}\right)+e_{33}\epsilon_{zz} 
\end{vmatrix}.
\end{equation}

In contrast to the case of planar quantum wells (QWs), the shear components of $\epsilon_{ij}$ must not be ignored for the NW geometry. We used the theoretical piezoelectric constants for GaN given by \citet{Shimada_jjap_2006} ($e_{33}=0.83$~C/m$^2$, $e_{31}=-0.45$~C/m$^2$, $e_{15}=-0.38$~C/m$^2$). These values seem appropriate also for (In,Ga)N, as the values of $e_{ij}$ for GaN and InN differ less than those for GaN between different theoretical and experimental studies.\cite{Shimada_jjap_2006} Figure~\ref{fig:FEM} shows that there is a considerable piezoelectric polarization in the (In,Ga)N insertions. In the center of the insertion, $|\vec{P}_{PZ}|$ amounts to 0.018~C/m$^2$ compared to 0.021~C/m$^2$ for a planar QW with the same thickness and composition profile. More importantly, even at the edge of the (In,Ga)N insertion, a value of 0.013~C/m$^2$ is observed. The elastic relaxation is actually not enhanced for an insertion extending to the NW sidewalls, and the piezoelectric polarization still dominates over the spontaneous polarization  whose contribution amounts to 0.005~C/m$^2$ already in the absence of strain.\cite{Fiorentini_apl_2002} A one-dimensional band profile evidencing the formation of strong internal electrostatic fields for the structure under investigation due to the piezoelectric polarization has been presented in Ref.~\onlinecite{Knelangen_nt_2011}. 

%=====================================================================
%%Fig.2
\begin{figure}
\includegraphics*[width=8cm]{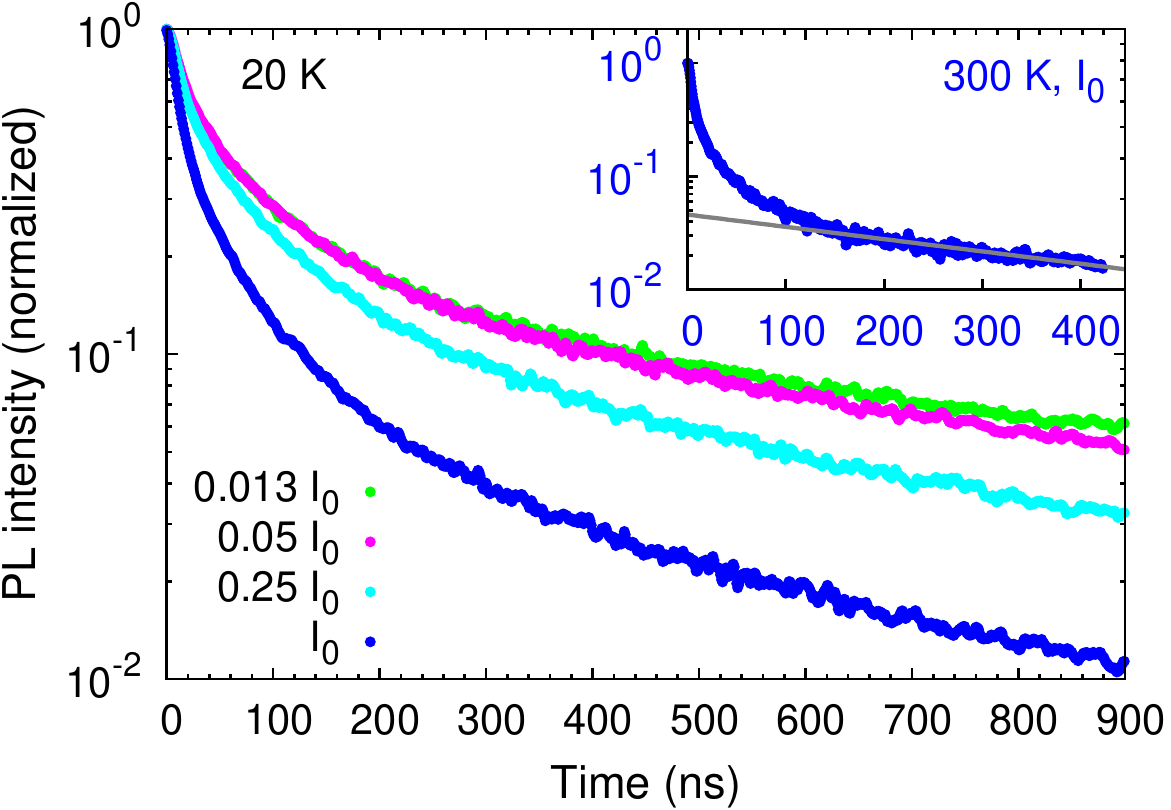}
\caption{\label{fig:TRPL}(Color online) PL transients of the spectrally integrated emission from the (In,Ga)N/GaN NW ensemble at 20~K and (inset) 300~K. Normalized transients for different excitation pulse intensities are plotted, where $I_0$ refers to the highest excitation density with $1.24\times10^{13}$~carriers/cm$^2$ per pulse. The long time asymptote at 300~K indicated by the gray line in the inset corresponds to a decay time of 400~ns.}
\end{figure}
%=====================================================================

%=====================================================================
%%Fig.3
\begin{figure}[b]
\includegraphics*[width=8cm]{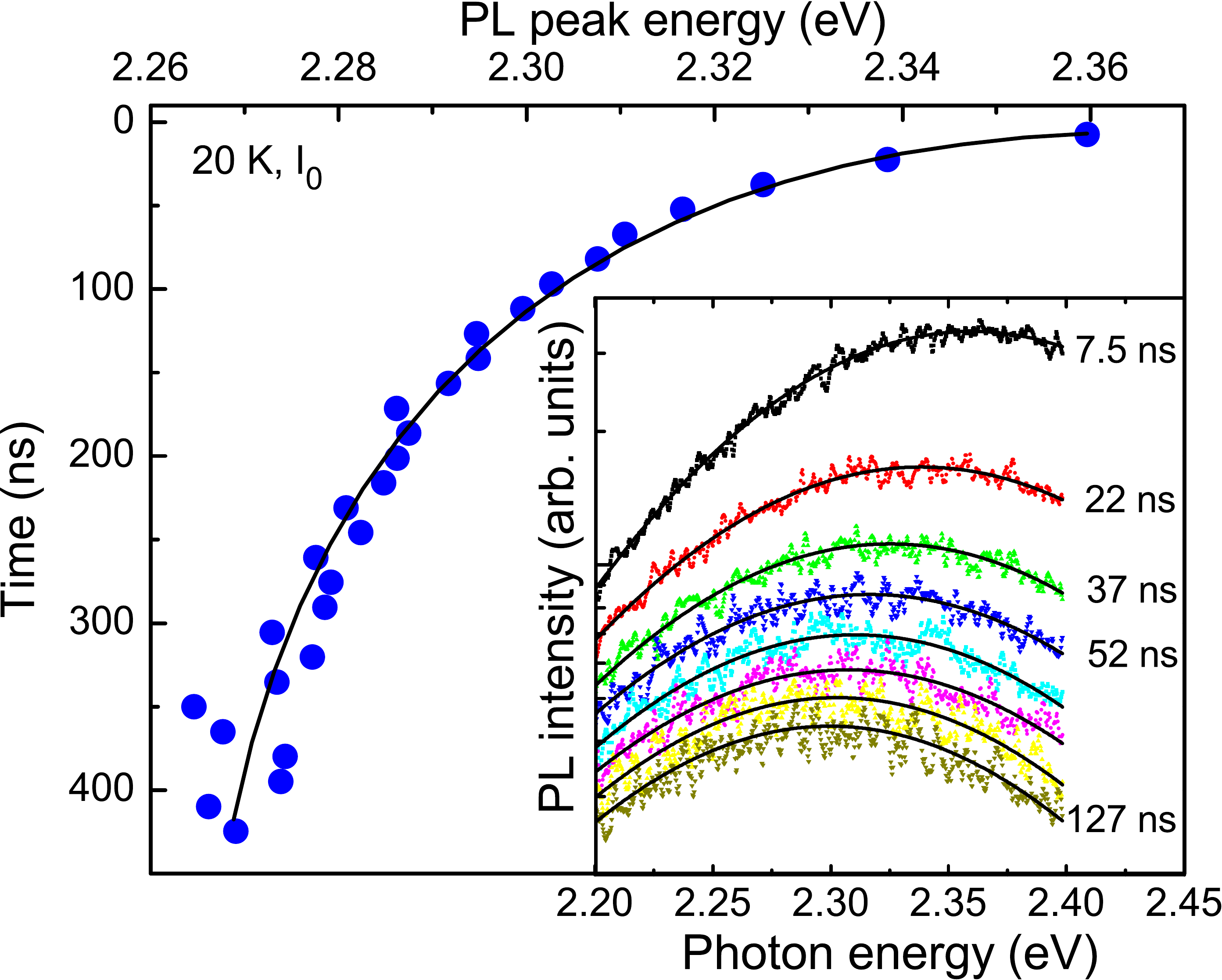}
\caption{\label{fig:TRPL2}(Color online) Evolution of the emission energy with decay time for the highest excitation density $I_0$ at 20~K. The line is a guide to the eye. The inset shows the corresponding transient spectra at different times as well as the Gaussian fits (solid lines) for obtaining the transition energies.}
\end{figure}
%=====================================================================

\subsection{Ensemble spectroscopy: TRPL}

Experimentally, the presence of piezoelectric fields in the NW ensemble under investigation should manifest itself in the 20~K PL transients presented in Fig.~\ref{fig:TRPL}. We first notice that the decay is rather slow, taking place over a timescale on the order of 1~$\mu$s. This is in contrast with measurements on (In,Ga)N/GaN NW heterostructures by \citet{Bardoux_prb_2009}, which yield an ensemble lifetime of $550$~ps for the slow component of a bi-exponential fit. As these authors observe luminescence from localization centers, our much longer lifetimes suggest a transition where the wave-function overlap is significantly reduced, consistent with our belief that the emission originates from a spatially indirect inter-well transition. For the present sample, the long decay time alone thus does not provide evidence for the presence of electric fields. However, we observe that the decay becomes faster with increasing excitation density (Fig.~\ref{fig:TRPL}). Such a behavior would not be expected for a transition being merely spatially indirect, but may very well be caused by a progressive screening of electrostatic fields. Note that the carrier sheet density giving rise to $|\vec{P}_{PZ}|=0.018$~C/m$^2$ is $n_s=1.13\times10^{13}$~cm$^{-2}$ and therefore close to the maximum excitation density $I_0$ (\emph{cf.}\ Fig.~\ref{fig:TRPL}). 

If our highest excitation screens the internal electrostatic fields, we would expect to observe a characteristic ``comma-shaped'' evolution of the peak energy with decay time.\cite{Lefebvre_prb_2004} As displayed in Fig.~\ref{fig:TRPL2}, this behavior is indeed observed: the emission band strongly red-shifts within the first 100~ns of the decay, followed by a much more gradual change. This strongly non-linear dependence of energy on time reflects the fact that the initial decrease of carrier density is fast because of the screening of the internal fields by the initially created carrier density. The decay of the carrier population progressively restores the internal fields, and the dynamics slows down correspondingly.\cite{Lefebvre_prb_2004} 

%=====================================================================
%%Fig.4
\begin{figure}
\includegraphics*[width=8cm]{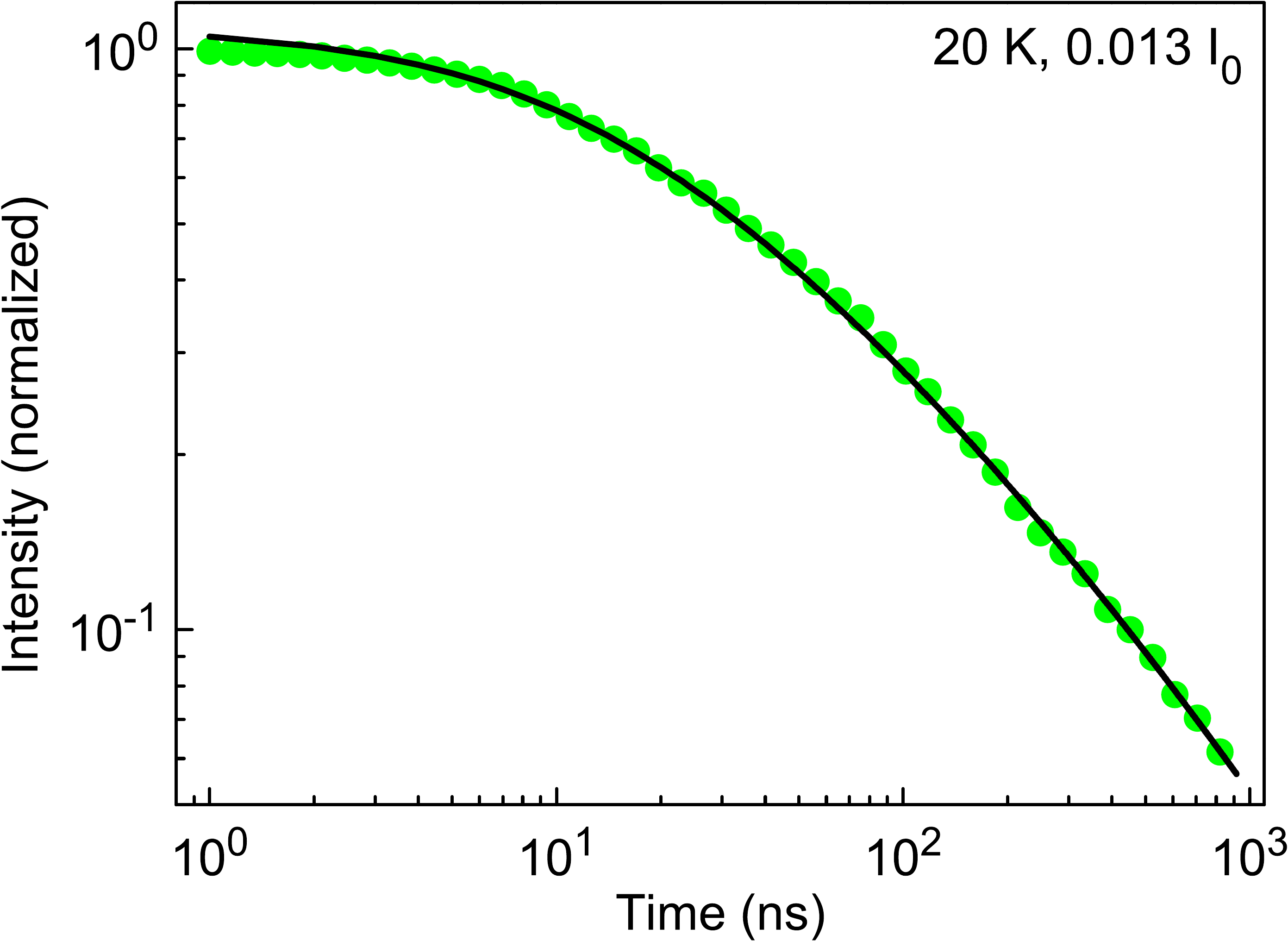}
\caption{\label{fig:TRPL3}(Color online) PL transient for the lowest excitation density $0.013 I_0$ at 20~K in double-logarithmic representation. For clarity, only a fraction of the data are displayed. The power-law decay is described  well by a donor-acceptor-pair recombination model (solid line).}
\end{figure} 
%=====================================================================

%=====================================================================
%%Fig.5
\begin{figure}[b]
\includegraphics*[width=8.4cm]{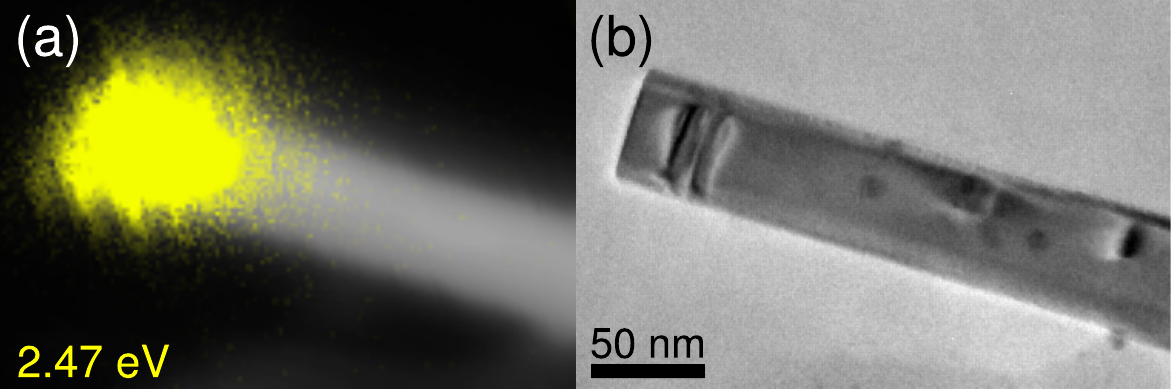} 
\caption{\label{fig:CL+TEM}(Color online) (a) Superposition of SEM and (false-color) monochromatic room-temperature CL images and (b) TEM image of the same NW dispersed on a TEM grid. The strain-induced contrast from the heterostructure coincides with the location of the light-emitting region.}
\end{figure}
%=====================================================================

At lower excitation densities, we still observe a spectral diffusion of the emission towards lower energies (not shown here), but far less pronounced than depicted above for the highest excitation. At the same time, the decay remains strongly non-exponential even at the lowest excitation density for which screening cannot possibly occur. Figure \ref{fig:TRPL3} displays this PL transient on a double-logarithmic scale. Clearly, the decay essentially follows a power-law. This dependence is characteristic for disordered systems in general\cite{Huntley_jpcm_2006} and for (In,Ga)N in particular.\cite{Morel_prb_2003, Brosseau_prb_2010} Regardless of the details, the interpretation of this peculiar decay type always relies on a spatial separation of charge carriers induced by the individual localization of electrons and holes in separate potential minima. Recombination is \emph{not} excitonic in this case, but occurs between electrons and holes with varying spatial separation. The solid line in Fig.~\ref{fig:TRPL3} is a fit to the data with the classical three-dimensional donor-acceptor-pair model of Thomas.\cite{Thomas_pr_1965} 

The low-temperature decay may thus be interpreted as follows. At low excitation densities, the broad emission band observed in the inset of Fig.~\ref{fig:TRPL2} stems from non-excitonic transitions between individually localized electrons and holes, resulting in a very slow, power-law decay of the emission intensity. These single particle states experience the vertical electrical field in the structure since they are spatially separated in vertical direction by the barrier between the two (In,Ga)N insertions. Their lateral separation will not influence the transition energy, but the overlap. Energy as well as overlap will depend sensitively on carrier density due to both a saturation of low-energy states and a screening of the internal fields at high excitation density.

Remarkably, the decay remains quite slow at elevated temperatures. The inset in Fig.~\ref{fig:TRPL} shows the PL transient obtained at room temperature. The long-time asymptote of the decay gives a lower bound of 400~ns for the non-radiative lifetime, significantly longer than values reported for state-of-the-art (In,Ga)N blue-to-green light-emitting and laser diodes.\cite{Li_apl_2010,Lutgen_pssa_2010} This result demonstrates the potential of (In,Ga)N/GaN NW heterostructures for applications in light emitting devices, and constitutes a major motivation to investigate the emission of this structure in more detail.

%=====================================================================
%%Fig.6
\begin{figure}
\includegraphics*[width=7.6cm]{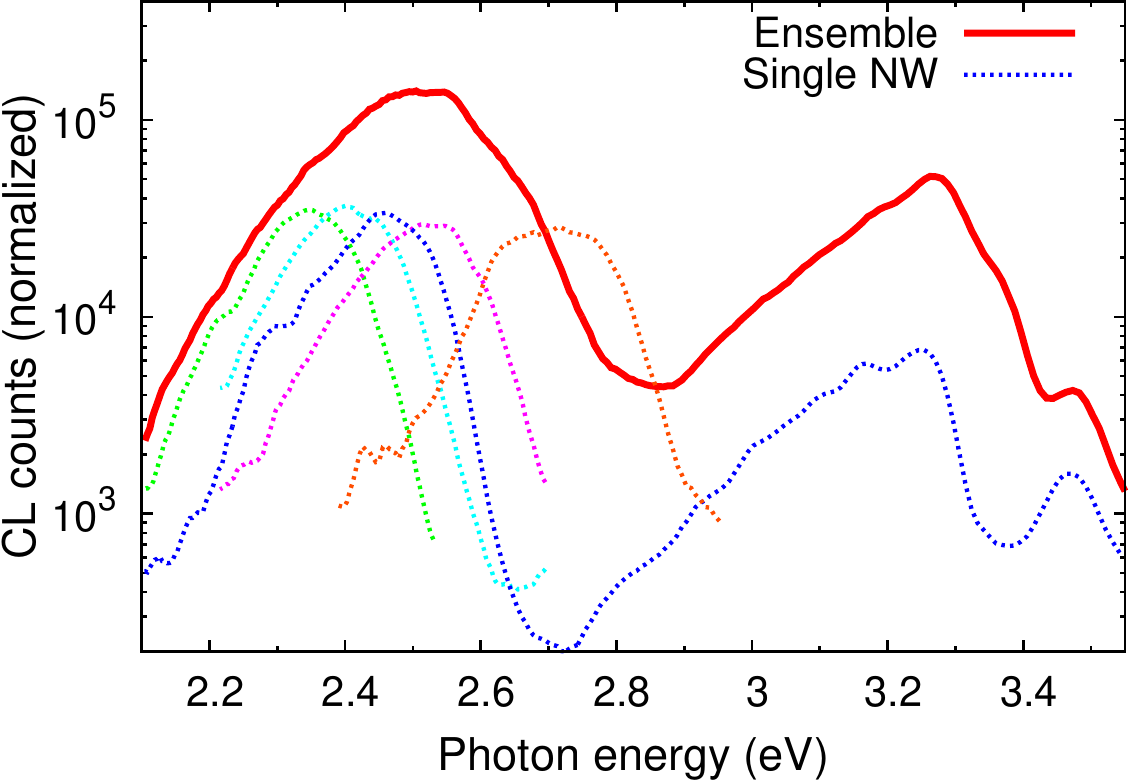} 
\caption{\label{fig:singleCL}(Color online) Comparison of normalized CL spectra from single NWs (dashed lines) and the NW ensemble (solid line) recorded at 10~K. The complete spectra exemplified for the ensemble and one NW exhibit additional contributions above 3~eV from the excitonic near-band edge luminescence and the donor-acceptor pair transition in the GaN NW base.}
\end{figure}
%===================================================================== 

\subsection{Single-wire spectroscopy: CL and $\mu$PL}

We start our investigation of single NWs with the successive imaging of the same single NW by SEM/CL and TEM as displayed in Figs.~\ref{fig:CL+TEM}(a) and \ref{fig:CL+TEM}(b), respectively. These experiments illustrate the spatial correlation of the green emission [hereafter referred to as the (In,Ga)N band] and the strain fields associated with the (In,Ga)N insertions. A complete CL spectrum measured at low temperature (10~K) on the NW ensemble is given in Fig.~\ref{fig:singleCL} (solid line) together with spectra from dispersed individual NWs (dashed lines). For the individual NWs, the emission energy of the (In,Ga)N band varies from NW to NW in the range of 2.3--2.7~eV. This proves that the (In,Ga)N band centered around 2.5~eV in the spectrum of the NW ensemble is indeed a superposition of these peaks. The individual bands remain rather broad with a full width at half maximum (FWHM) around 160~meV compared to 300~meV for the NW ensemble. This considerable line width is consistent with the existence of a near-continuum of localized states in line with our interpretation of the decay dynamics above. Variations of the mean In content of a few percent and of the barrier thickness (2--3~nm) observed between different NWs in TEM (not shown here) readily account for the spread of the emission energy over 400~meV.

%=====================================================================
%%Fig.7
\begin{figure}
\includegraphics*[width=7.6cm]{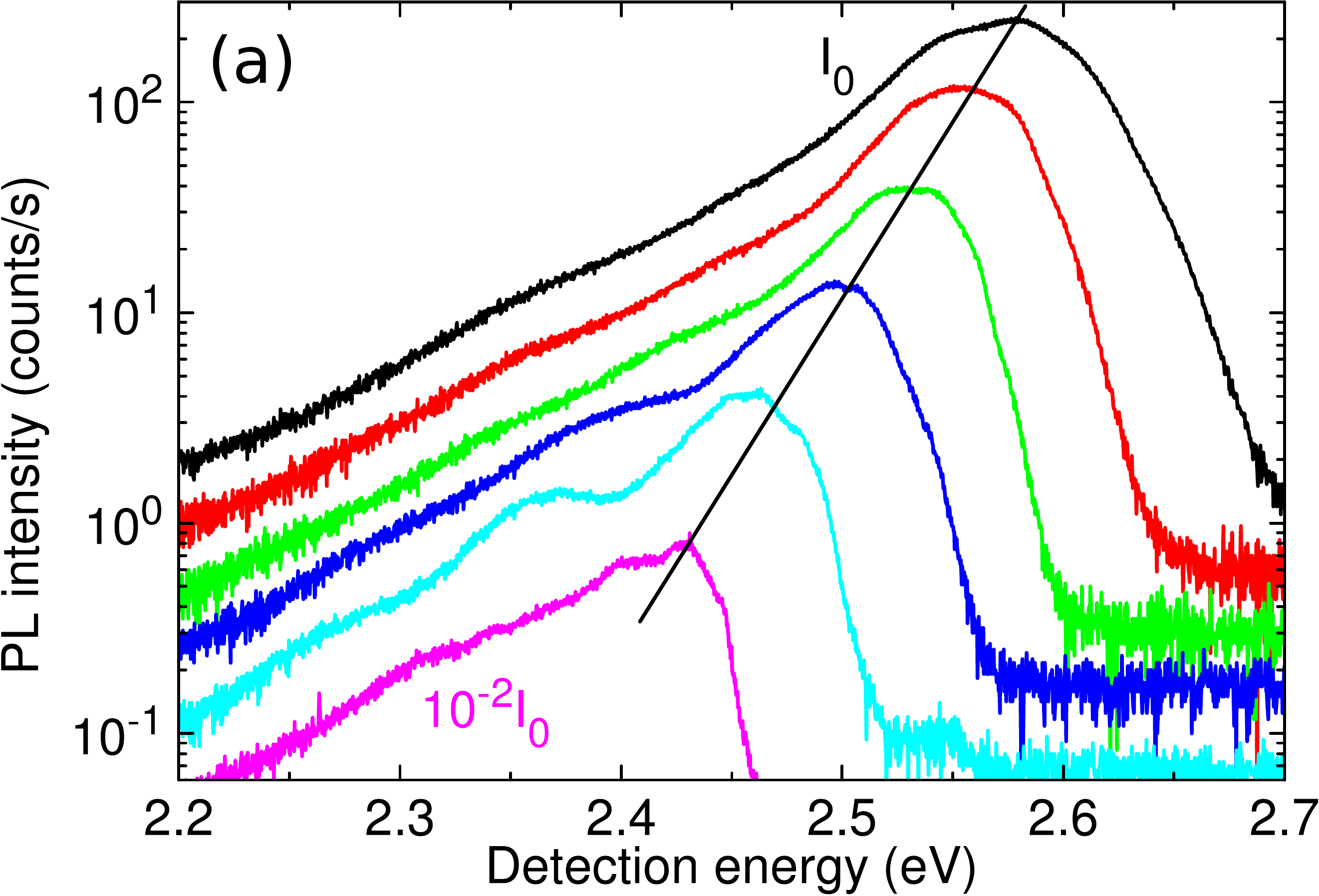}\\[2mm] 
\includegraphics*[width=7.6cm]{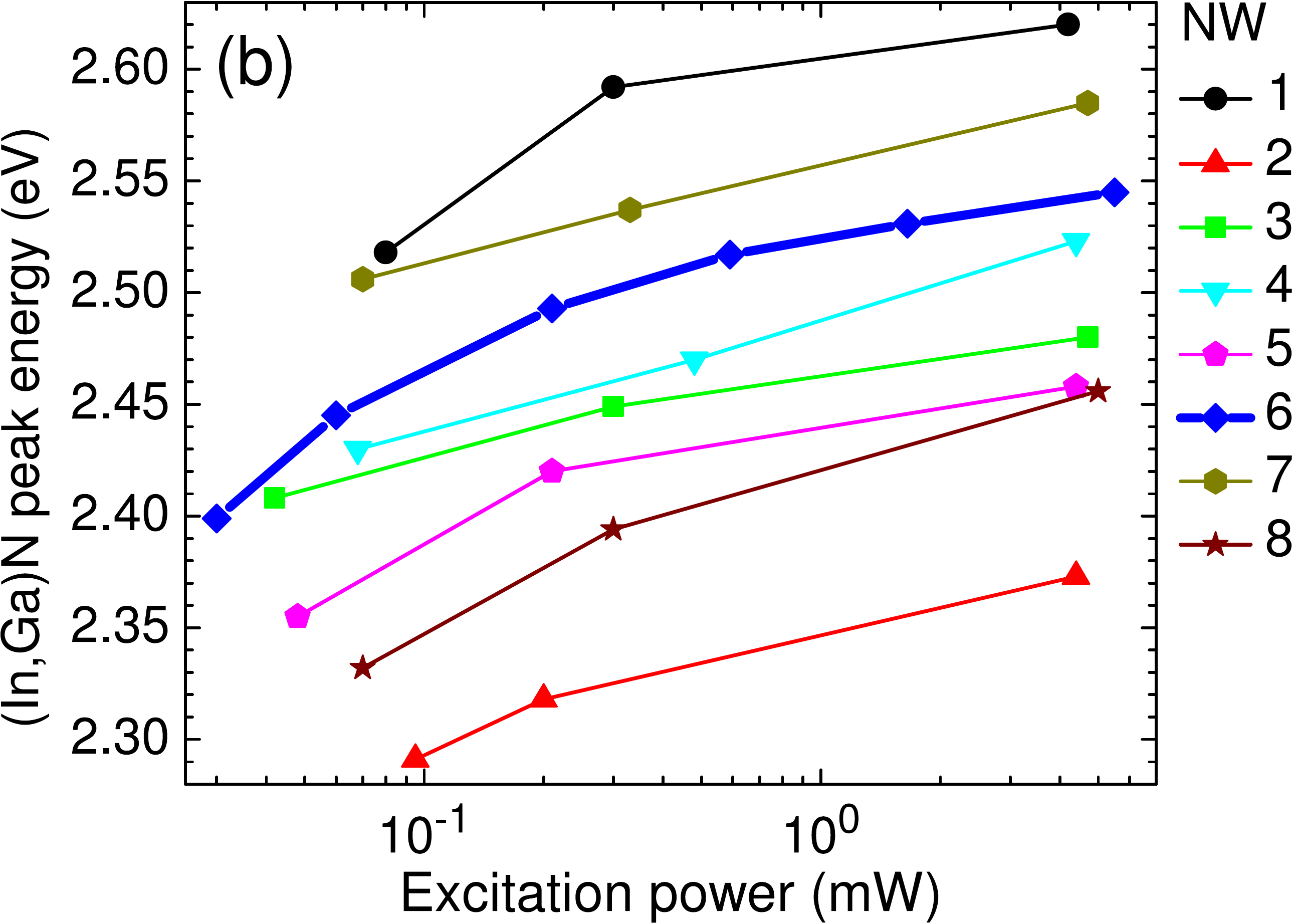}
\caption{\label{fig:PLsingle}(Color online) (a) $\mu$PL spectra for an individual NW  with the excitation density increasing from bottom to top and (b) peak energy versus excitation power (logarithmic scale) of the (In,Ga)N band for different NWs with the NW displayed in (a) highlighted by the thick line.}
\end{figure}
%=====================================================================

To improve the spectral resolution and to reduce the excitation density, we performed $\mu$PL at 10~K on single dispersed NWs according to the procedure described in Ref.~\onlinecite{Brandt_prb_2010}. Figure~\ref{fig:PLsingle}(a) shows the $\mu$PL spectra for an individual NW taken with excitation densities spanning two orders of magnitude. Over the entire range of excitation densities, the integrated intensity exhibits a strictly linear dependence on excitation density. At the same time, the (In,Ga)N band exhibits a significant blue-shift with increasing excitation density. Note that the spectra are basically \emph{rigidly} blue-shifted, with little symmetric broadening and no asymmetric broadening towards higher energies except for the highest excitation density. Also note that the low-energy tails of the spectra are parallel to each other. These findings rule out band filling as the cause of the blue-shift, render the saturation of low-energy localized states unlikely, and thus leave the progressive screening of internal electrostatic fields as the most likely cause of the blue-shift observed.\cite{Peng_apl_1999, Kuokstis_apl_2002, Holmes_apl_2009} To further confirm this conclusion, we performed power-dependent measurements for eight different single NWs as shown in Fig.~\ref{fig:PLsingle}(b). The magnitude of the peak shift is very similar for all investigated NWs. Looking at the more detailed measurement highlighted in this semi-logarithmic plot, it is clear that the dependence is weaker than logarithmic. This peculiar dependence follows the behavior predicted by \citet{Kuroda_jap_2002} and by \citet{Pinos_apl_2008} for free carrier screening of the QCSE. The saturation of the peak shift with excitation density simply arises from the fact that the lifetime decreases once screening is effective, thus requiring disproportionately higher excitation densities to further increase the carrier density. Contrary to expectation,\cite{Pinos_apl_2008} we do not observe a saturation of the transition energy for low excitation densities. We believe this finding to be a consequence of the very long carrier lifetime and the resulting high carrier densities for cw excitation. At the same time, this high carrier density even for our lowest excitation density is probably also responsible for the lack of any fine-structure in the spectra due to individual localized states, which we would otherwise expect to emerge in single-wire spectroscopy. 

%=====================================================================
%%Fig.8
\begin{figure}[t!]
\includegraphics*[width=7.8cm]{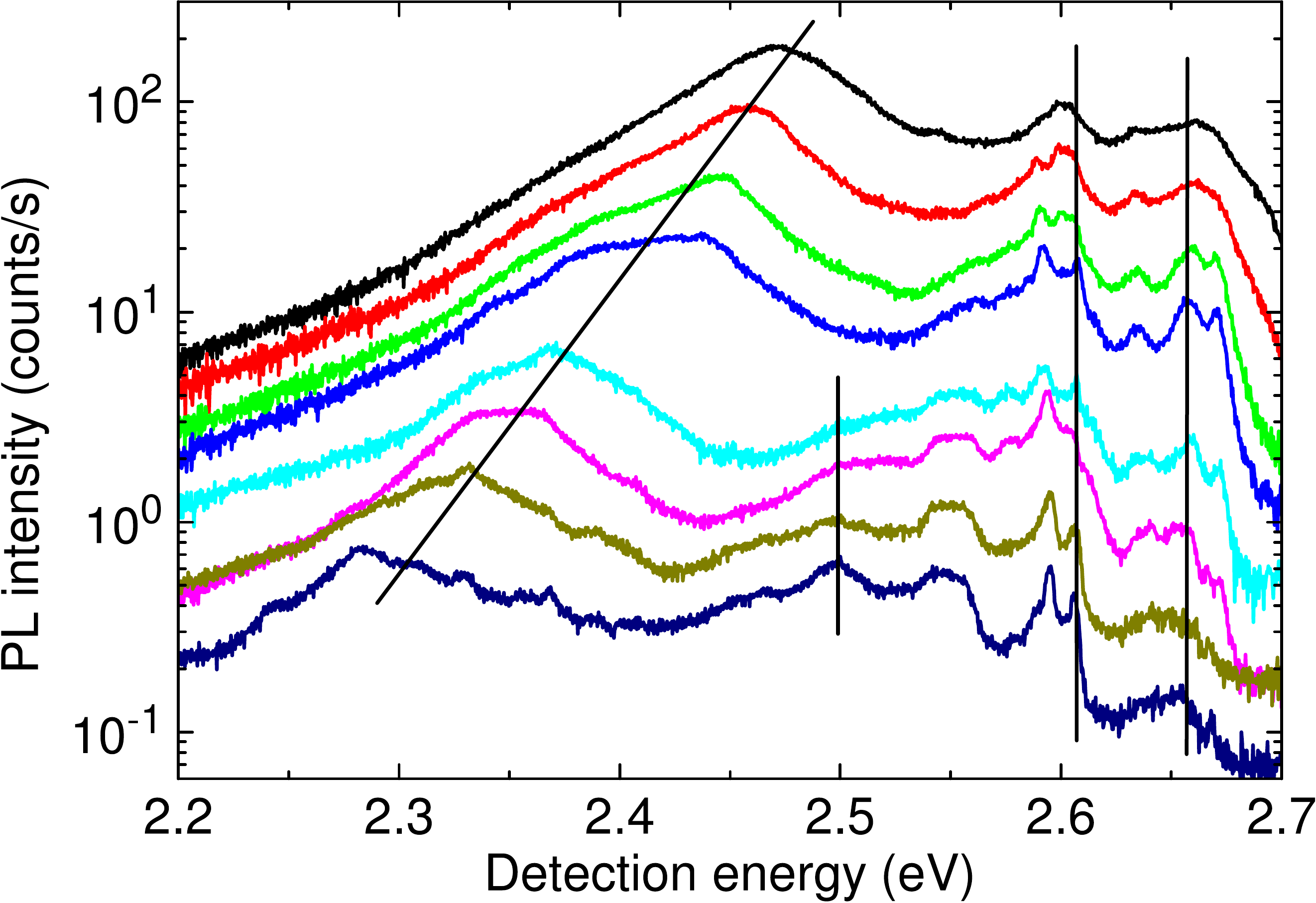}
\caption{\label{fig:PLsingle2}(Color online) PL spectra for a NW that exhibits additional sharp lines at an energy higher than that of the main (In,Ga)N band. These lines do not shift with excitation power (increasing from bottom to top by approximately two orders of magnitude).}
\end{figure}
%===================================================================== 

However, besides the broad emission band visible on the low energy side of the spectrum, additional sharp PL peaks at a slightly higher energy appear in the spectra of many NWs. An example of such a spectrum is shown in Fig.~\ref{fig:PLsingle2}. The positions of these sharp PL peaks are insensitive to the excitation density as emphasized by the vertical lines. This fact and their small FWHM of down to 3--6~meV are fingerprints of zero-dimensional localized states. In fact, localized states originating from compositional fluctuations are known to manifest themselves in spectrally narrow transitions in planar (In,Ga)N/GaN QWs when measured using nano-apertures\cite{Schomig_prl_2004} as well as in (In,Ga)N/GaN NW heterostructures.\cite{Bardoux_prb_2009} 

Provided that the spatial extent of the associated potential minima is on the order of one nm, the resulting localized states facilitate spatially direct transitions and are thus not affected by the polarization field. Note that these direct transitions are related to localized \emph{excitons}, in contrast to the individually localized electrons and holes discussed above. Taking into account the strong confinement in the nm-scale potential minima, the energy of the corresponding localized states would be only slightly below that of the surrounding (In,Ga)N matrix, which was measured to have an In content of 0.2. According to the In$_x$Ga$_{1-x}$N band gap (neglecting excitonic effects) of
\begin{equation}
  E_\mathrm{G}(x)=3.51(1-x)+0.69x-1.72x(1-x)
\end{equation}
with the parameters taken from Ref.~\onlinecite{Schley_prb_2007} corrected for temperature, this In content corresponds to an energy of 2.67~eV, i.e., right at the high energy cutoff of the sharp lines visible in Fig.~\ref{fig:PLsingle2}. We note that, while all investigated NWs show the broad emission band, a few do not exhibit the additional sharp lines [\emph{cf.} Fig.~\ref{fig:PLsingle}(a)]. In other words, the inter-well transition appears in all NWs, while strong exciton localization within one insertion occurs in most but not all NWs.

\section{Conclusion}

Our results show that both the QCSE and localization play an important role in the (In,Ga)N/GaN NW heterostructure under investigation. We believe, however, that our results are of significance for (In,Ga)N/GaN NW heterostructures in general. In fact, (In,Ga)N insertions of about 10~nm thickness being embedded within the GaN NW are evident from the TEM images presented in at least two other studies.\cite{Chang_apl_2010,Armitage_nt_2010} Our FEM simulations reveal the elastic relaxation of such structures to be marginal, causing strong piezoelectric fields to reside within the insertions. The associated drastic decrease of the electron-hole overlap renders a regular QW transition unlikely. In fact, the green (In,Ga)N band seen in the present work is a result of an \emph{inter-well} transition due to the two closely spaced insertions, enabling us to experimentally observe the QCSE. Moreover, three-dimensional band-structure calculations for (In,Ga)N segments in NWs presented recently by \citet{Boecklin_prb_2010} conclusively show that the elastic relaxation may even result in a complete \emph{lateral} separation of electrons and holes, inhibiting recombination entirely. In light of these complications, emission from strongly localized states is a likely candidate for the luminescence observed from these structures. In particular, we have shown in the present work that emission energy of strongly localized excitons may be entirely insensitive to excitation density even in the presence of a strong electrostatic field. The absence of an excitation-dependent blue-shift observed by other groups\cite{Armitage_nt_2010,Lin_apl_2010,Nguyen_nl_2011} thus may be due to localization and not to an absence of polarization fields.\\

\section*{Acknowledgement}

We would like to thank Martin Wölz for a critical reading of the manuscript. The growth of the NWs was partially funded by the German BMBF joint research project \textsc{Mona\-Lisa} (Contract No.\ 01BL0810).

\end{document}